\documentclass[%
reprint, 
nofootinbib,
 amsmath,amssymb,
 aps, physrev,
]{revtex4-2}

\usepackage{graphicx}
\usepackage{dcolumn}
\usepackage{bm}
\usepackage[colorlinks=true,linkcolor=blue,urlcolor=blue,citecolor=blue]{hyperref}

\newcommand{\wphicdm}{$w_{\phi}$CDM}

\newcommand{\refcite}[1]{Ref.~\cite{{#1}}}

\newcommand{\aap}{A\&A}

\newcommand{\aj}{A.J.}
\newcommand{\mnras}{MNRAS}

\newcommand{\jcap}{JCAP}
\newcommand{\apjl}{ApJL}
\newcommand{\pasp}{PASP}

\newcommand{\refedit}[1]{{{#1}}}

\begin{document}


\title{\refedit{Scalar field} dark energy models: Current and forecast constraints}

\author{Anowar J.~Shajib$^{1,2,3,}$}\email{ajshajib@uchicago.edu}\thanks{NHFP Einstein Fellow} \author{Joshua A. Frieman$^{1,2,4,}$}\email{jfrieman@uchicago.edu}%
 
\affiliation{%
$^1$Department of Astronomy \& Astrophysics, University of Chicago, Chicago, IL 60637, USA \\
$^2$Kavli Institute for Cosmological Physics, University of Chicago, Chicago, IL 60637, USA \\
$^3$Center for Astronomy, Space Science and Astrophysics, Independent University, Bangladesh, Dhaka 1229, Bangladesh \\
$^4$SLAC National Accelerator Laboratory, 2575 Sand Hill Road, Menlo Park, CA 94025
}
%

\date{\today}

\begin{abstract}
Recent results from Type Ia supernovae (SNe Ia) and baryon acoustic oscillations (BAO), in combination with cosmic microwave background (CMB) measurements, have focused renewed attention on dark energy models with a time-varying equation-of-state parameter, $w(z)$. In this paper, we describe the simplest, physically motivated models of evolving dark energy that are consistent with the recent data, a broad subclass of the so-called thawing scalar field models \refedit{that we dub \wphicdm}. We provide a quasi-universal, quasi-one-parameter functional fit to the scalar-field $w_\phi(z)$ that captures the behavior of these models more informatively than the standard $w_0w_a$ phenomenological parametrization; their behavior is completely described by the current value of the equation-of-state parameter, $w_0=w(z=0)$. Combining current data from BAO (\refedit{DESI Data Release 2}), the CMB (\textit{Planck} and ACT), large-scale structure (DES Year-3 $3\times2$pt), SNe Ia (DES-SN5YR), and strong lensing (TDCOSMO + SLACS), \refedit{for \wphicdm} we obtain \refedit{$w_0=-0.904_{-0.033}^{+0.034}$, 2.9$\sigma$} discrepant from the $\Lambda$ cold dark matter ($\Lambda$CDM) model. The Bayesian evidence ratio substantially favors this \wphicdm\ model over $\Lambda$CDM. The data combination that yields the strongest discrepancy with $\Lambda$CDM is BAO+SNe Ia, for which  \refedit{$w_0=-0.837^{+0.044}_{-0.045}$, $3.6\sigma$} discrepant from $\Lambda$CDM and with a Bayesian evidence ratio strongly in favor. We find that the so-called $S_8$ tension between the CMB and large-scale structure is slightly reduced in these models, while the Hubble tension is slightly increased. We forecast constraints on these models from near-future surveys (DESI-extension and the Vera C.~Rubin Observatory LSST), showing that the current best-fit \wphicdm\ model will be distinguishable from $\Lambda$CDM at over 9$\sigma$.
\end{abstract}

\maketitle


\section{Introduction}\label{sec:intro}

Over the last quarter century, the $\Lambda$ cold dark matter ($\Lambda$CDM) model, with 70\% vacuum energy (equivalently the cosmological constant $\Lambda$) and 30\% matter (CDM plus baryons), has become the standard cosmological paradigm \cite[e.g.,][]{Frieman2008}.  Until recently, it has proven consistent with an array of increasingly precise measurements of cosmic structure and expansion history \cite[e.g.,][]{PlanckCollaboration18,Alam21,Abbott22}. Yet it has long been recognized that $\Lambda$CDM is just the simplest version of a broader class of models that can include more complex behaviors in the dark matter or dark energy sectors. 

Interest in {\it dynamical} dark energy models \cite{Frieman95} arose in the mid-1990s from what appeared to be observational challenges to the emerging $\Lambda$CDM paradigm from strong lensing statistics \cite{Kochanek96}, the high cosmic microwave background (CMB) normalization of the matter power spectrum \cite{Coble97}, and early Type Ia supernova (SN Ia) results \cite{Perlmutter96}. While those challenges were subsequently resolved by new data \cite{Riess98, Perlmutter99} or improved data modeling \cite{Chiba99, Keeton02, Mitchell05}, or both, the most recent results from large SN Ia samples such as Pantheon+ \cite{Brout22}, Union \cite{Rubin23}, and the Dark Energy Survey (DES) \cite{ DESCollaboration24}, and baryon acoustic oscillation (BAO) measurements from the Sloan Digital Sky Survey (SDSS) \cite{Alam21} and the Dark Energy Spectroscopic Instrument (DESI) \cite{DESI24}, in combination with \textit{Planck} CMB anisotropy constraints \cite{PlanckCollaboration18}, are again suggesting that evolving dark energy models may provide a better fit to the data than $\Lambda$CDM. 

Neglecting spatial fluctuations, in the fluid approximation, the impact of dark energy on cosmological observables can be quantified in terms of its equation-of-state parameter, the ratio of its energy density to its pressure, $w=p/\rho$. An often-used, two-parameter phenomenological prescription for the redshift evolution of $w$ is given by $w(z)=w_0+w_a z/(1+z)$ \cite{Chevallier01, Linder03}, where $w_0$ is the current value, $w_a$ characterizes its time evolution, and $z$ is redshift; $\Lambda$CDM corresponds to the particular case $w_0=-1$ and $w_a=0$. The Pantheon+ SN compilation \cite{Brout22}, in combination with the SDSS BAO and \textit{Planck} CMB results, yielded $w_0=-0.841^{+0.066}_{-0.061}$, $w_a=-0.65^{+0.28}_{-0.32}$ for a spatially flat Universe (which we assume throughout), consistent with $\Lambda$CDM at 2$\sigma$. The DES 5-year SN survey data (DES-SN5YR) \cite{DESCollaboration24}, a much larger high-redshift SN sample than previous ones, when combined with the SDSS BAO and \textit{Planck} CMB results and DES Year-3 weak lensing and galaxy clustering \cite{Abbott22}, yielded   $w_0=-0.773^{+0.075}_{-0.067}$, $w_a=-0.83^{+0.33}_{-0.42}$, roughly 3$\sigma$ from $\Lambda$CDM. Subsequently, the DESI collaboration \cite{DESI24} released its first-year BAO results \refedit{(DESI DR1)} and combined them with \textit{Planck} CMB and with three separate SN samples. Using the DES-SN5YR sample, they found the greatest (3.9$\sigma$) tension with $\Lambda$CDM, $w_0=-0.727\pm 0.067$, $w_a=-1.05^{+0.31}_{-0.27}$. When using the combination of DESI \refedit{DR1} and SDSS BAO data with the most precise BAO distance measurement in each redshift bin, they obtained a combined constraint $w_0=-0.761\pm 0.064$, $w_a=-0.88^{+0.29}_{-0.25}$, 3.5$\sigma$ from $\Lambda$CDM and quite close to the DES-SN5YR results in \refcite{DESCollaboration24}. \refedit{The subsequent combination of DESI DR2 BAO data, \textit{Planck} CMB, and DES-SN5YR yielded very similar but more precise constraints, $w_0=-0.752\pm0.057$, $w_a=-0.86^{+0.23}_{-0.20}$ \cite{DESIDR2}, 4.2$\sigma$ discrepant from $\Lambda$CDM.}

While convenient for separating ``thawing" ($w_a<0$) \cite{Frieman95, Coble97} from ``freezing" ($w_a>0$) \cite{Ratra88, Steinhardt99} dark energy models \cite{Caldwell05} (the latter are now disfavored by the data) and for comparing the constraining power of different probes \cite{Albrecht06}, the $w_0w_a$ parametrization used in these and previous analyses is not ideal for expressing constraints on physically motivated dark energy models. For $w_0+w_a<-1$, which includes essentially all of the allowed parameter space at 3$\sigma$ in the recent DESI BAO+CMB+SN analysis, the null energy condition (NEC), $w(z)>-1$, is {\it naively} violated above some redshift, and the best-fit parameter region corresponds to NEC violation above $z = 0.35$. Put another way, a flat prior on $w_0$ and $w_a$ does not restrict consideration to physically motivated models, which is problematic given the relatively weak constraining power of current data in this parameter space. 

An important caveat to the statement in the previous paragraph attaches to the word ``naively". Since the function constrained by observations is the expansion history, $H(z)$, and integrals thereof, not $w(z)$ itself, and since the inverse mapping from $H(z)$ and distances $D(z)$ to $w(z)$ is not unique, it is in principle possible to find apparently NEC-violating values of $w_0$ and $w_a$ that closely reproduce the expansion history of an NEC-preserving, thawing scalar-field model with equation of state parameter $w_\phi(z) \geq -1$, even though $w_\phi(z)$ itself is not (necessarily) well fit by the resulting $w_0w_a$ model \cite{Linder03,dePutter08, Linder08, Lodha24, Shlivko24}. While such a mapping is possible, it means that we do not have a physical interpretation of the parameters in the $w_0w_a$ prescription. \refedit{Moreover, models with dark matter-dark energy interactions can yield an {\it effective} $w<-1$ while preserving the NEC \cite{Agrawal2021, Khoury2025}.}

In this paper, we review the theoretical motivations for \refedit{a class of} physics-inspired dark energy models consistent with the recent data -- what have come to be called thawing scalar field models -- and explore present and future constraints upon them, without the mediation of the $w_0w_a$ parametrization. The simplest versions of these models \refedit{-- in terms of the conventional form of the scalar field potential $V(\phi)$ --} exhibit a quasi-universal behavior \cite{Linder08},   \refedit{and we label these collectively the \wphicdm\ model.} In Section \ref{sec:theory}, we provide a one-parameter fit to the evolution of the scalar field equation-of-state parameter, $w_\phi(z)$, that is extremely accurate for a wide class of these models \cite[as mentioned in Ref.][]{Camilleri24} \refedit{(see also \cite{Shlivko2025})}. In Section \ref{sec:constraints}, we constrain this parametrized model using a broad array of currently available datasets. We then forecast expected constraints from near-future surveys in Section \ref{sec:forecasts}, showing the range of model parameters that will be distinguishable at high confidence from $\Lambda$CDM in the coming years. Lastly, we conclude the paper in Section \ref{sec:conclusion}.

\section{Theory}\label{sec:theory}

As noted above, early interest in dynamical dark energy models was driven in part by observations that appeared to disfavor $\Lambda$CDM but also by theoretical considerations, in particular the lack of a fundamental explanation for the very small (relative to the Planck scale $M_{\rm Pl}=1/\sqrt{G}=1.2\times 10^{19}$ GeV, where $G$ is the gravitational constant) but non-zero value of the vacuum energy density, $\rho_{\rm vac} \sim 10^{-120}M_{\rm Pl}^4$, within a factor of 2--3 of the matter density today. 

A potential solution is to postulate that the true vacuum energy density of the Universe is, in fact, zero (perhaps due to a symmetry) and that the current non-zero {\it effective} vacuum energy is due to a cosmologically evolving field that has not yet reached its ground state. At the current epoch, any such field must be extremely light by particle physics standards, with effective mass $m \alt 3H_0 \simeq 4.5\times 10^{-33}$ eV. This idea was inspired by models of primordial inflation, 
in which a massive scalar field takes much longer than a Hubble time to reach its ground state, giving rise to an effective vacuum energy density that drives an early epoch of accelerated expansion.

For an ultra-light scalar field $\phi$ to serve as the source of dark energy, its current energy density must satisfy $\rho_\phi(t_0)=\Omega_{\phi} \rho_{\rm crit}=3\Omega_{\phi} H_0^2 M_{\rm Pl}^2/8\pi$, where $\Omega_{\phi} \simeq 0.7$ is the current vacuum energy density parameter, and its kinetic energy must be small compared to its potential energy, that is, $\dot{\phi}^2/2 \ll V(\phi)$. For a free, massive scalar field with potential energy density $V(\phi)=m^2\phi^2/2$, the constraint above on $m$ implies that the amplitude of the field must be close to the Planck scale, $\phi \agt M_{\rm Pl}/10$, so that $m/\phi \sim 10^{-60}$ or smaller.
These are generic features of scalar-field dark energy models. Such an extreme hierarchy of scales is not natural in most scalar quantum field theories. If the scalar field interacts with other fields, one expects quantum corrections to generate a much larger mass and self-coupling for the field. For example, both the mass and amplitude (vacuum expectation value) of the Higgs field of the Standard Model are of order the electroweak scale, $\sim$100 GeV. 

Pseudo--Nambu--Goldstone bosons (pNGBs), of which the QCD axion is an example, which arise in models with spontaneously broken global symmetries, can naturally exhibit such a large hierarchy, since their small mass is protected from quantum corrections by a residual shift symmetry that is only weakly broken, and their interactions with other fields are suppressed by inverse powers of the large spontaneous symmetry breaking scale $f$. Their potential energy density is generically of the form
\begin{equation}
V(\phi) = m^2 f^2 \left[1+\cos\left(\frac{\phi}{f}\right)\right]~,
\label{eq:pngb}
\end{equation}
where $m$ is the pNGB effective mass. For such a field to provide the dark energy, the discussion above implies $f \agt 10^{18}$ GeV, and the shift-symmetry breaking scale $M \equiv \sqrt{mf} \simeq 3\times 10^{-3}$ eV \cite{Frieman95}. In the recent literature, these models are often referred to as ultra-light axions. For comparison, for the QCD axion, $M \simeq 100$ MeV, typically $f \sim 10^{12}-10^{16}$ GeV for axion dark matter, and the axion mass $m=M^2/f$ is of order $10^{-5}-10^{-9}$ eV.

Assuming the canonical Lagrangian for a scalar field, ${\cal{L}}= (1/2)g^{\mu\nu}\partial_\mu \phi \partial_\nu \phi -V(\phi)$, neglecting spatial perturbations, the equation of motion of the field in an expanding universe is given by
\begin{equation} \label{eq:phiEOM}
\ddot{\phi}+3H\dot{\phi}+\frac{\mathrm{d}V}{\mathrm{d}\phi}=0~,
\end{equation}
where the expansion rate at late times is given by
\begin{equation}
H^2 = \frac{8\pi}{3M_{\rm Pl}^2}\left(\rho_{\rm m} +\rho_\phi\right)~.
\end{equation}
Here, $\rho_{\rm m}$ is the matter density, and the energy density of the homogeneous scalar field is 
\begin{equation}
\rho_\phi = \frac{1}{2}\dot{\phi}^2+V(\phi)~.
\end{equation}
The time-evolution of $\rho_\phi$ is determined by $H$ and by the equation-of-state parameter, $w_\phi=p_\phi/\rho_\phi$, where the scalar-field pressure is 
\begin{equation}
p_\phi=\frac{1}{2}\dot{\phi}^2-V(\phi)~.
\end{equation}
For a given form of the potential, $V(\phi)$, and initial value of the scalar field, $\phi(t_{\rm i}) \equiv \phi_{\rm i}$ at some early time $t_{\rm i} \ll t_0$, this dynamical system can be solved to obtain $\phi(t)$ and thus the expansion history (assuming spatial flatness) as
\begin{equation}
\frac{H(z)^2}{H^2_0} = E(z)^2 =\Omega_{\rm m} (1+z)^{3}+ \Omega_{\phi}\, g_\phi(z)~,
\end{equation}
where $g_\phi(z) \equiv \rho_{\phi} (z) / \rho_{\phi}(z=0)$ is the redshift-dependent scaling of the scalar field energy density.

In ``thawing'' scalar field models,  the driving term $\mathrm{d}V/\mathrm{d}\phi$ in Eq.(\ref{eq:phiEOM}) is subdominant at early times compared to the Hubble-damping term $3H\dot{\phi}$. In this limit, the field is effectively frozen at its initial value $\phi_{\rm i}$, hence $\dot{\phi}(t_{\rm i})=0$, $\rho_\phi(t_{\rm i})=V(\phi_{\rm i})$, and $w_\phi(t_{\rm i})=-1$. Once the Hubble expansion rate drops below the curvature of the potential, $H(z) < \sqrt{|\mathrm{d}^2V/\mathrm{d}\phi^2|}\sim m$, the field begins to roll, develops non-negligible kinetic energy, and $w_\phi$ grows from $-1$ \cite[e.g.,][]{Rosenfeld07}. Thawing models include standard potentials of the form $V(\phi)=\pm(1/2)m^2 \phi^2 + \lambda \phi^4$ (with $\lambda>0$), the pNGB model above, and polynomials $V(\phi)=\sum_{i=1}^{n}a_i\phi^i$ with $a_i \geq 0$. 
Note that the ``thawing'' evolution of $w_\phi$ corresponds to $w_a<0$ in the $w_0w_a$ parametrization \cite{Caldwell05}, the region of parameter space favored by the recent data. 

As an example, for a free, massive scalar with $V(\phi)=(1/2)m^2 \phi^2$, the condition $\Omega_{\phi}=0.7$ implies $(m/H_0)\,\phi(t_0)/M_{\rm Pl} \simeq 0.44$ in the limit where $\dot{\phi}^2 \ll V(\phi)$. For $m/H_0 \agt 1$ (or $\alt 1$), the field begins rolling before the present epoch (or not), and the present value of the equation-of-state parameter $w_0 \equiv w_\phi(t_0)$ can be measurably above $-1$ (or not). For the massive scalar model with $\Omega_{\phi}=0.7$, numerically we find $w_0 \simeq -1+(1/7)(m/H_0)^2$ to good approximation \refedit{for $m \alt H_0$.}

\begin{figure*}
    \centering
    \includegraphics[width=\textwidth]{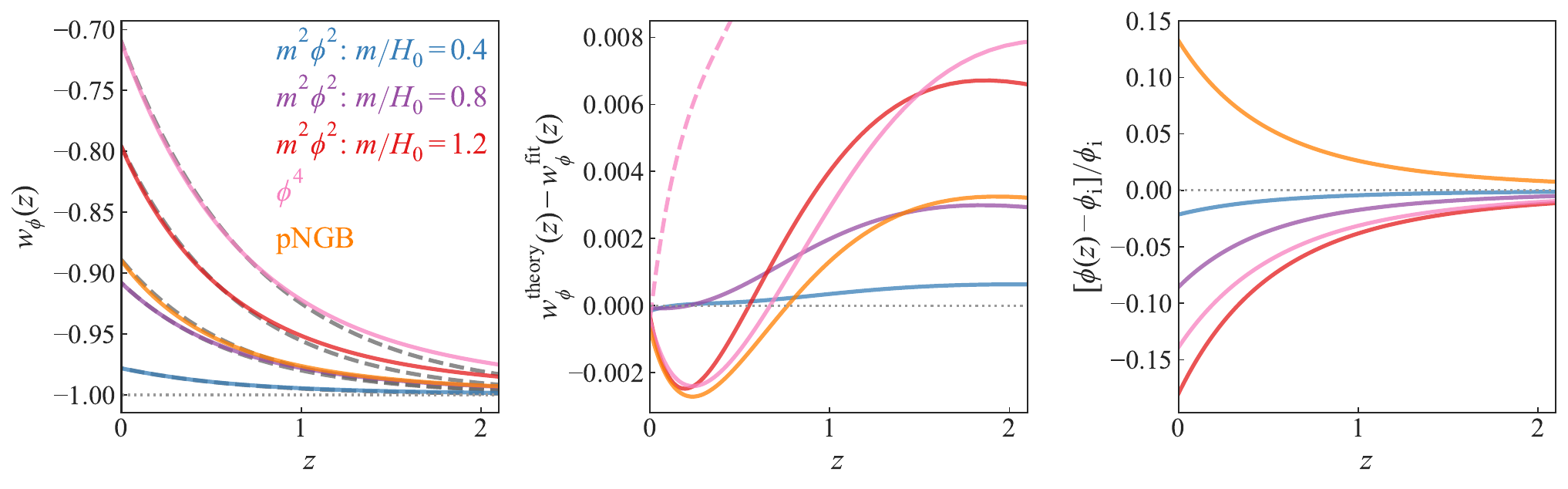}
    \caption{\label{fig:evol}
    \textit{Left panel:}  $w_\phi(z)$ for thawing scalar field models with $\Omega_{\phi} \simeq 0.7$: massive scalar with $m/H_0=0.4 ~(\phi_{\rm i}/M_{\rm Pl}=1.1)$ (blue), $0.8 ~(0.55)$ (purple), $1.2 ~(0.4)$ (red), $V=(1/2)m^2 M_{\rm Pl}^2 (\phi/M_{\rm Pl})^4$ model with $1.2 ~(0.65)$ (pink), and pNGB model with $f=0.25M_{\rm Pl}, ~m/H_0=1.2, ~\phi_{\rm i}/f=1.5$ (orange). Dashed grey curves show the corresponding fits $w^{\rm fit}_{\phi}(z)$ (Eq. \ref{eq:walpha}) with $(w_0, \alpha)= (-0.710, 1.35), (-0.796, 1.52), (-0.889, 1.6), (-0.908, 1.52), (-0.978, 1.45)$, from upper to lower. \textit{Middle panel:} residuals between the $w_\phi(z)$ of scalar models and the $w^{\rm fit}_{\phi}(z)$  fits for the same models; for comparison, dashed pink curve shows the (larger) residual for the $\phi^4$ model using the constrained $w_a=-1.58(1+w_0)$ model of \refcite{Lodha24} (see text) instead of $w^{\rm fit}_\phi$. \textit{Right panel:} scalar field evolution for these model parameters.
    }
\end{figure*}

The evolution of several scalar field models is shown in Fig.~\ref{fig:evol}. 
While there have been a variety of approximate solutions and fits to late-time scalar field evolution in the literature \cite[e.g.,][]{Linder08, Dutta08, dePutter08, Chiba09}, numerical experiments show that the redshift-evolution of $w_\phi$ for these simple thawing models is very well approximated by the form \cite{Camilleri24}
\begin{equation} \label{eq:walpha}
w^{\rm fit}_\phi(z)=-1+(1+w_0)e^{-\alpha z}~,
\end{equation}
which has the asymptotic behaviors $w^{\rm fit}_\phi(z=0)=w_0$, $w^{\rm fit}_\phi(z\gg1/\alpha)=-1$. The best-fit value of $\alpha$, \refedit{which sets the characteristic redshift interval $\Delta z \sim 1/\alpha$ over which $w(z)$ evolves,} is only weakly dependent on $w_0$ and on the form of $V(\phi)$ and is generally in the narrow range $\alpha=1.35-1.55$; this \refedit{value and} narrow range reflects the fact that in these models the evolution timescale is approximately set by the scalar field mass, which is of order $H_0$ as noted above. For all the cases shown in Fig.~\ref{fig:evol}, this function fits the scalar-field $w_\phi(z)$ to better than 0.004 (0.008) at redshifts $z <1$ ($z<2$) (middle panel of Fig.~\ref{fig:evol}), fits the theoretical luminosity distance at $z<2$ to a fractional error better than 0.04\%, and fits the scalar-field expansion history $H(z)$ to better than 0.05\% over the same redshift range, well below the precision of current and future measurements. The approximation of Eq. \eqref{eq:walpha} holds if the effective scalar mass $\sqrt{|\mathrm{d}^2V/\mathrm{d}\phi^2|}$ is not large compared to $H_0$; otherwise, the field will begin oscillating around the minimum of its potential before the present epoch and will not give rise to cosmic acceleration at late times. At the other limit, $\sqrt{|\mathrm{d}^2V/\mathrm{d}\phi^2|} \ll H_0$, as noted above, the field will still be frozen, $w_0=-1$, and the model is indistinguishable from $\Lambda$CDM. 

From Eq. \eqref{eq:walpha}, the redshift-scaling of the scalar field energy density is given by
\begin{equation} \label{eq:g_fit_exact}
    g_\phi^{\rm fit} (z) = \exp \left[3 e^\alpha \, (1 + w_0) \left\{{E_1}(\alpha) - E_1(\alpha+\alpha z) \right\} \right]~,
\end{equation}
where ${E_1}(x)$ is the exponential integral. The exponential integral $E_1(x)$ has a fast computational algorithm through the convergent series
\begin{equation}
    E_1(x) = -\gamma - \ln x - \sum_{k=1}^\infty \frac{(-x)^k}{k\,k!}~,
\end{equation}
where $\gamma \approx 0.577216$ is the Euler--Mascheroni constant \cite{Abramowitz64}. Numerical implementations of $E_1(x)$ are also readily available in widely used mathematical libraries such as the \textsc{GNU} Scientific Library (\textsc{gsl}) and \textsc{scipy}. 

Note that if we were to treat Eq. \ref{eq:walpha} as a purely phenomenological model with two free parameters, then $\alpha=0$ would correspond to a $w$CDM model with constant $w^{\rm fit}_\phi(z)=w_0$, while $\alpha \gg 1$ would correspond effectively to $\Lambda$CDM, with $w^{\rm fit}_\phi(z)=-1$ down to very low redshift and a sharp transition to $w^{\rm fit}_\phi(0)=w_0$ at $z \sim 1/\alpha \ll 1$. However, these extreme values of $\alpha$ are not realized in the family of simple scalar field models we are considering. There are thawing scalar models that do not fall into this family; e.g., the hill-top potential \cite{Dutta08} with very steep fall-off, $V(\phi)=M^4[1-(1/2)k^2(\phi/M_{\rm Pl})^2]$, with $k^2 \gg 1$ \cite{Shlivko24}, is \refedit{only} roughly fit by a model of the form of Eq. \ref{eq:walpha} with larger values of $\alpha$ than we consider here. We do not consider such models, because (a) they are {\it not} an approximation to the natural pNGB potential (Eq. \ref{eq:pngb}) near its maximum, for which $k \sim 1$ (which is included in our model class), and (b) as noted in \refcite{Shlivko24}, they require fine-tuning of initial conditions to give rise to accelerated expansion that is observably distinguishable from $\Lambda$CDM, with initial field values $\phi_{\rm i} \sim M_{\rm Pl}/k^2 \ll M_{\rm Pl}$. The current data provide no meaningful constraints on the parameter $\alpha$ if it is left free; in Sec. \ref{sec:forecasts}, we comment on the potential to constrain $\alpha$ with future data.

\refedit{For comparison with previous literature, another parametrization of thawing models we briefly} consider is the $w_0w_a$ model fit with the constraint (fixed prior) \cite{Linder08, Lodha24}: $w_a=-1.58(1+w_0)$. This fit has a low-$z$ redshift-dependence similar to that of Eq. \ref{eq:walpha} with $\alpha=1.35-1.55$ and provides a good fit to $H(z)$ for thawing scalar models (though a less accurate fit to $w_\phi(z)$, see middle panel of Fig.~\ref{fig:evol} for an example). However, it does not satisfy the asymptotic limit $w(z) \rightarrow -1$ as $z \rightarrow \infty$, \refedit{which is a defining feature of thawing models and is built into the form of Eq. \ref{eq:walpha}; in fact, in this $w_0w_a$ model $w(z)<-1$ for $z>1.72$.} 

\section{Current constraints}\label{sec:constraints}

In this section, we obtain current cosmological constraints on thawing scalar field models from different cosmological probes using the $w^{\rm fit}_\phi$ formula in Eq. \ref{eq:walpha} with flat prior on $\alpha \sim \mathcal{U}( 1.35, 1.55)$, which we refer to hereafter as the  \wphicdm\ model. The datasets we use are BAO (Section \ref{sec:bao_data}), CMB (Section \ref{sec:cmb_data}), large-scale structure (LSS, Section \ref{sec:lss_data}), SNe Ia (Section \ref{sec:sn_data}), and strong lensing time delays (Section \ref{sec:sl_data}). We present the constraints from these datasets in Section \ref{sec:result}. 

\subsection{BAO dataset}\label{sec:bao_data}

\refedit{We use the BAO dataset from DESI Data Release 2 (DR2) \citep{DESIDR2}, which provides measurements of $D_{\rm M}/r_{\rm d}$ and $D_{\rm H}/r_{\rm d}$ or $D_{\rm V}/r_{\rm d}$ in seven redshift bins with $z_{\rm eff}=$ 0.295, 0.510, 0.706, 0.934, 1.321, 1.484, and 2.330.}
Here, $r_{\rm d}$ is the sound horizon at the drag epoch, $D_{\rm M}(z)$ is the comoving distance to redshift $z$, $D_{\rm H}(z) \equiv c/H(z)$ is the Hubble distance, and $D_{\rm V}(z) \equiv \left[z\,D^2_{\rm M}(z)\,D_{\rm H}(z)\right]^{1/3}$. 

\refedit{To investigate the gain in precision and potential shifts in the parameter values between DESI DR1 and DR2, we also consider for comparison the BAO distance ladder combining SDSS and DESI DR1 BAO measurements, as prescribed by \refcite{DESI24}.}
In this combined distance ladder, there are eight redshift bins with $z_{\rm eff}=$ 0.15, 0.38, 0.51, 0.706, 0.93, 1.317, 1.491, and 2.33. For the three bins at $z_{\rm eff}<0.6$, the SDSS measurements \cite{Alam21} are chosen due to their higher measurement precision than DESI DR1 in this redshift range. For the Ly$\alpha$ BAO at $z_{\rm eff}=2.33$, the measurements from SDSS and DESI DR1 are combined. For the four redshift bins in between, the DESI DR1 measurements are chosen since they have the highest precision.
    
\subsection{CMB dataset} \label{sec:cmb_data}

We combine the primary CMB power spectra from \textit{Planck} \cite{PlanckCollaboration20} with the CMB lensing datasets from \textit{Planck} and the Atacama Cosmology Telescope (ACT) DR6 \cite{Madhavacheril24, Qu23}. Specifically, we use the high-$\ell$ $TT$, $TE$, and $EE$ power spectra through the \texttt{Plik\_lite} likelihood, low-$\ell$ ($\ell \leq 30$) $TT$ power spectra using the \texttt{Commander} likelihood, and low-$\ell$ $TE$ power spectra using the \texttt{simall} likelihood from the official PR3 release (i.e., TT$+$lowE in \textit{Planck} notation). For the CMB lensing dataset, \textit{Planck}'s \textsc{NPIPE} PR4 \cite{Carron22} is combined with ACT DR6, as implemented in the likelihood module released with the latter (i.e., the \texttt{actplanck\_baseline} variation of the lensing power spectra). To achieve the minimum settings (or higher) recommended by ACT for numerical accuracy in \textsc{camb} \cite{Howlett12, Lewis00}, we set \texttt{lmax = 4000}, \texttt{lens\_margin = 1250}, \texttt{lens\_potential\_accuracy = 4}, \texttt{AccuracyBoost = 1.1}, \texttt{lSampleBoost = 1}, \texttt{lAccuracyBoost = 1}. For the \textsc{halofit} module in \textsc{camb}, we use \textsc{hmcode-2020} \cite{Mead21} excluding baryonic feedback.
    
\subsection{LSS dataset} \label{sec:lss_data}

We use the three two-point ($3 \times 2$pt) correlation functions of LSS probes from the DES Year-3 dataset \cite{Abbott22}: (i) autocorrelation of cosmic shear from weak lensing in four source-galaxy photometric-redshift bins, (ii) autocorrelation of lens-galaxy clustering in six photo-$z$ bins, and (iii) cross-correlation of source-galaxy shear with lens-galaxy positions (i.e., galaxy--galaxy lensing). For the lens-galaxy sample, we use the MagLim sample that was adopted as the baseline in \refcite{Abbott22}. When inferring the cosmological parameters from this dataset, we marginalize over 25 nuisance parameters, and the adopted priors on them are identical to those in \refcite{Abbott22}.

\subsection{SN Ia dataset} \label{sec:sn_data}

\refedit{For the SNe Ia data, we use the DES-SN5YR dataset \citep{DESCollaboration24}, which is by far the largest, single-survey, high-redshift SNe Ia dataset to date and incorporates a number of analysis improvements over the Pantheon+ analysis, as described in  \cite{Vincenzi2025}.} \refedit{By comparison, the high-redshift SN data in the Pantheon+ and Union data sets are based on smaller compilations of SN data from multiple surveys with different photometric calibration systems.} 
\refedit{As described in \refcite{Vincenzi24}, the high-redshift DES-SN5YR dataset, with distance moduli $\mu_i(z)$ for 1,635 photometrically classified SNe Ia in the redshift range $0.10 < z < 1.13$, is combined with an external, low-redshift sample of 194 SNe Ia at $0.025 < z < 0.10$. This low-$z$ SNe Ia data set is a high-quality subset of several low-redshift samples, chosen to minimize systematic errors and take advantage of the photometric uniformity of the high-redshift sample. By contrast, Pantheon+ and Union used larger, more heterogeneous, low-redshift samples with more varied data quality. As a result, the combined DES-SN5YR analysis should have a greater control over systematic errors than the other SN compilations.}

\subsection{Strong lensing dataset} \label{sec:sl_data}

We utilize the latest compiled likelihood from the Time-Delay Cosmography (TDCOSMO) collaboration \cite{Birrer20}, which incorporates strong lensing time delay information from seven lensed quasars \cite{Suyu10, Suyu14, Wong17, Birrer19b, Chen19, Jee19, Rusu20, Wong20, Shajib20}. We choose to include additional constraints on the stellar orbital anisotropy from nine Sloan Lens ACS \cite[SLACS;][]{Bolton06} survey lenses with integral-field unit kinematics \cite{Czoske12} (i.e., the TDCOSMO$+$SLACS\textsubscript{IFU} dataset in the notation of \refcite{Birrer20}; see \refcite{Hogg24} for usage of the same dataset, with additional priors and extensions, to constrain the $w$CDM and $w_0w_a$CDM cosmologies). We marginalize the nuisance parameters pertaining to the mass-sheet and mass-anisotropy degeneracies by simultaneously sampling them with the cosmological parameters.

\newcommand{\uu}{\mathcal{U}}
\newcommand{\nn}{\mathcal{N}}

\begin{table}
	\centering
	\caption{
        \label{tab:priors}
        Adopted priors on the cosmological and nuisance parameters relevant to the various datasets used in this paper. See \refcite{PlanckCollaboration18} for definitions of the cosmological parameters that are not defined in this paper and of the CMB parameters, and \refcite{Birrer20} for definitions of the strong lensing parameters. The stated uniform priors for $\Omega_b,..., \sum m_\nu$ are only used when LSS or CMB datasets are used. For the other datasets, we fix these four parameters at the best-fit \textit{Planck} 2018 values \cite{PlanckCollaboration18}, since the other probes are not sensitive to them.
        }
    \renewcommand{\arraystretch}{1.3}
    \begin{ruledtabular}
    \begin{tabular}{ll}
    Parameter & Prior \\
    \hline
    \textbf{Cosmology} & \\
    $h$ & $\uu(0.55, 0.91)$ \\
    $\Omega_{\rm m}$ & $\uu (0.1, 0.9)$ \\
    $w_0$ & $\uu(-1, -0.33)$ or $\uu(-2, -0.33$) \\
    $\alpha$ & $\uu(1.35, 1.55)$ \\
    
    
    $\Omega_{\rm b}$ & $\uu(0.03, 0.07)$ \\
    $n_{\rm s}$ & $\uu(0.87, 1.07)$ \\
    $A_{\rm s}\times10^{9}$ & $\uu (0.5, 5)$ \\
    $\sum m_{\nu}$ [eV] & $\uu (0.06, 0.6)$ \\
    $N_{\rm eff}$ & $\delta(3.046)$ \\
    
    \textbf{BAO} & \\
    $r_{\rm d}h$ [Mpc], instead of $h$ & $\uu (10, 1000)$ \\
		
	\textbf{CMB} & \\
	$A_{\it Planck}$    & $\nn (1, 0.0025^2)$ \\
	$A_{\rm Lens}$ & $\delta(1)$ \\
        $\tau$ & $\mathcal{U}(0.01, 0.25)$ \\
    
	\textbf{LSS} & \\
	25 parameters & Same ones from \refcite{Abbott22} \\
    
	\textbf{SNe Ia} & \\
    Absolute magnitude, $M$ & $\uu (-21, -18)$ \\
    
	\textbf{Strong lensing} & \\
	$\lambda_{\rm MST}$ & $\uu (0.5, 1.5)$ \\
	$\log \sigma_{\lambda_{\rm MST}}$ & $\uu (-3, -0.3)$ \\
	$\alpha_{\lambda}$ & $\uu (-1, 1)$ \\
	$\log a_{\rm ani}$ & $\uu (-1, 0.7)$ \\
	$\log \sigma_{a_{\rm ani}}$ & $\uu (-2, 0)$ \\
	
    \end{tabular}
    \end{ruledtabular}	
    \renewcommand{\arraystretch}{1}
\end{table}

\subsection{Current results} \label{sec:result}

We use the \textsc{CosmoSIS} software program \cite{Zuntz15} to implement and combine the likelihoods from the above datasets and sample from the posterior using the \textsc{nautilus} software package \citep{Lange23}, which can also compute the Bayesian evidence through importance nested sampling. To set the background cosmology, we modify \textsc{camb}, including its \textsc{halofit} module, to implement the \wphicdm\ model. We list our adopted priors for all cosmological and other nuisance parameters in Table \ref{tab:priors}. We show results for two different flat, bounded priors on the parameter $w_0$: $w_0 \sim \mathcal{U}(-2, -0.33)$ and $w_0 \sim \mathcal{U}(-1, -0.33)$. The latter explicitly imposes the NEC $w(z) \geq -1$, which is satisfied by the canonical scalar field models as discussed above. In all cases, we assume the Universe is spatially flat.

\begin{figure}
    \centering
    \includegraphics[width=\columnwidth]{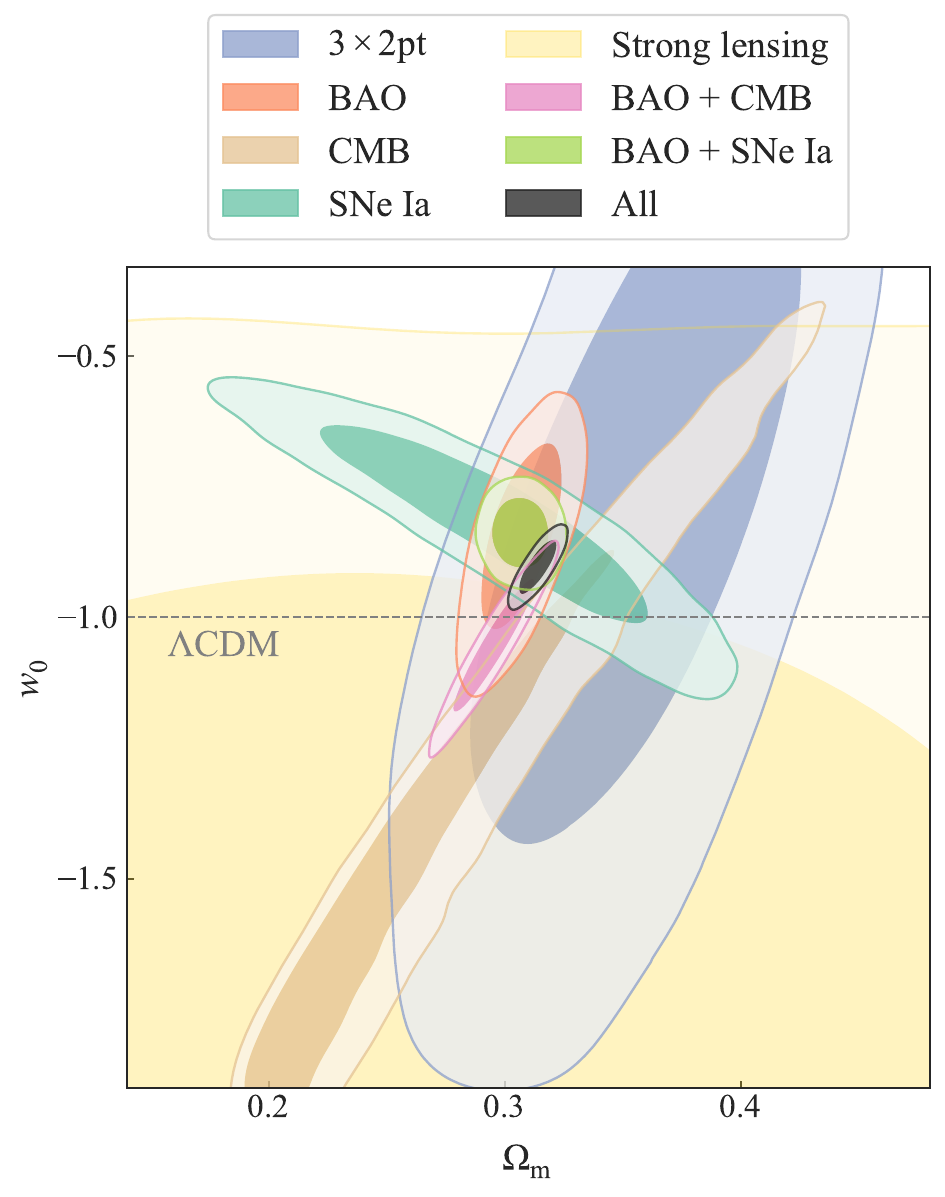}
    \caption{Current constraints on  $\Omega_{\rm m}$ and $w_0$ in the \wphicdm\ cosmology from various cosmological datasets and some of their combinations, assuming the broader prior on $w_0$. \refedit{The utilized datasets are: DES Year-3 for the $3\times2$ pt, DESI DR2 for BAO, \textit{Planck}+ACT for CMB, DES-SN5YR for SNe Ia, and TDCOSMO for strong lensing.} The darker and lighter shaded regions for each probe trace the 68\% and 95\% credible regions, respectively. The horizontal dashed grey line indicates the $\Lambda$CDM model with $w_0=-1$.}
    \label{fig:omegam_w_constraints}
\end{figure}

\begin{figure}
    \centering
    \includegraphics[width=\columnwidth]{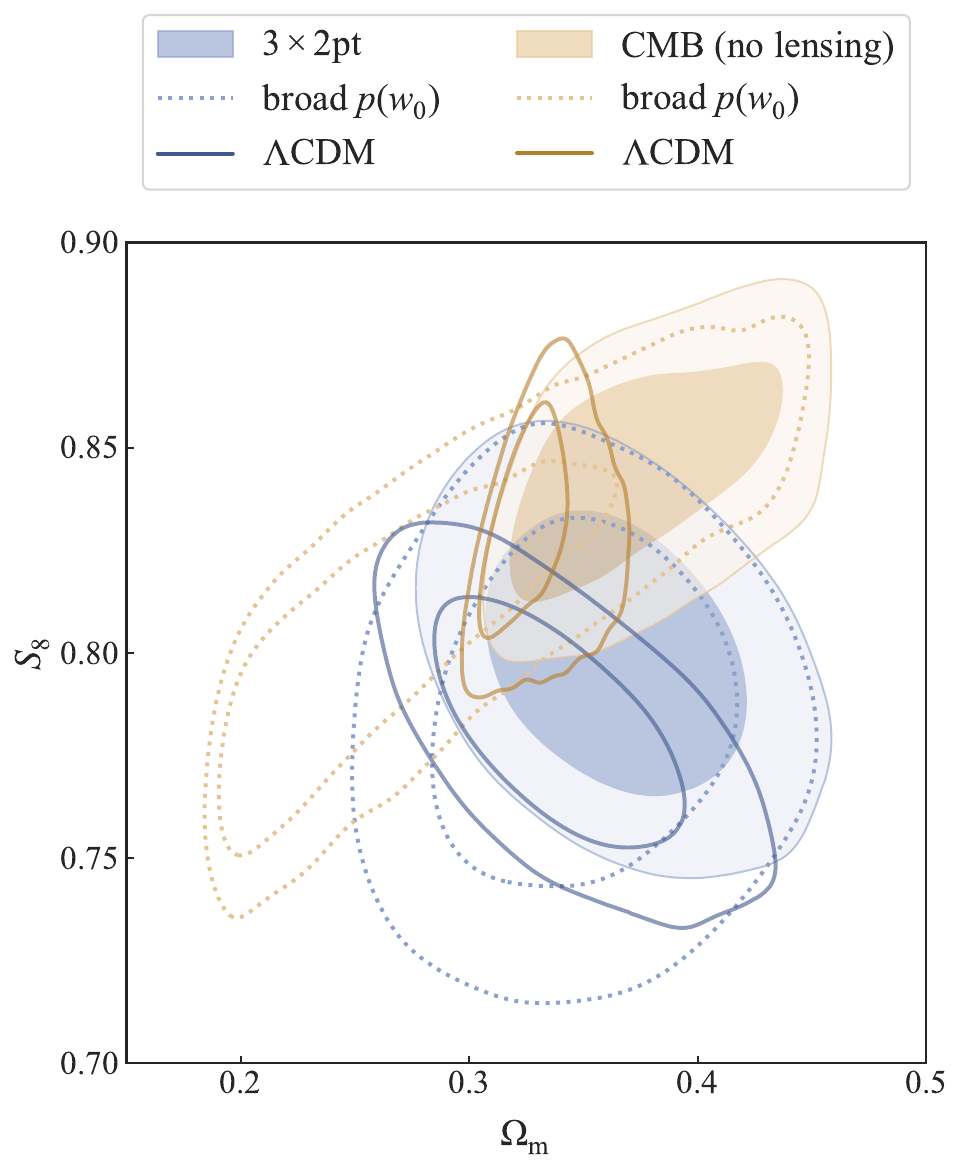}
    \caption{\label{fig:s8_constraints}
    Current constraints from DES Year-3 $3\times2$pt and \textit{Planck} CMB (no lensing) on  $\Omega_{\rm m}$ and $S_8 \equiv \sigma_8 \left( \Omega_{\rm m} / 0.3 \right)^{0.5}$ 
    in three cosmologies: (i) the \wphicdm\ model with the narrow NEC prior on $w_0$ (shaded contours), (ii) the \wphicdm\ model with the broad prior on $w_0$ (dotted contours), and (iii) the $\Lambda$CDM model (solid contours). The inner and outer contours represent the 68\% and 95\% credible regions, respectively. Slight ($<0.3\sigma$) differences from  Fig.~16 of \refcite{Abbott22} for $\Lambda$CDM are likely due to our use of an improved model for the non-linear matter power spectra from \refcite{Mead21} in the \textsc{halofit} module instead of the one from \refcite{Takahashi12}.
    }
\end{figure}

The resulting constraints on the Hubble constant $h=H_0/(100\ \text{km}\ \text{s}^{-1} \text{Mpc}^{-1})$, the matter density parameter $\Omega_{\rm m}$, the dark energy equation-of-state parameter $w_0$ (Eq. \ref{eq:walpha}), and the density perturbation amplitude $S_8=\sigma_8(\Omega_{\rm m}/0.3)^{0.5}$ are shown in Table \ref{tab:cosmo_params} and Figs. \ref{fig:omegam_w_constraints} and \ref{fig:s8_constraints}. In Table \ref{tab:cosmo_params}, we also give the Bayesian evidence ratio between the \wphicdm\ and $\Lambda$CDM models (plus priors). In practice, the parameter constraints are insensitive to the precise width and shape of the prior on $\alpha$, provided it is sufficiently narrow (the data have no constraining power on $\alpha$, because observables vary negligibly as $\alpha$ is varied over its prior range for fixed $w_0$). 

\begin{table*}
    \caption{
        \label{tab:cosmo_params}
        Cosmological parameter constraints on \wphicdm\ from different datasets. The first two rows for each dataset correspond to the two alternative priors on $w_0$: $\mathcal{U}(-2, -0.33)$ and $\mathcal{U}(-1, -0.33)$, while the third corresponds to $\Lambda$CDM (i.e., $w_0  = -1$). The provided point estimates are medians, and the $1\sigma$ uncertainties are obtained from the 16th and 84th percentiles. The last column provides the Bayesian evidence ratio between the \wphicdm\ model and the $\Lambda$CDM model (plus priors); a value of $\log_{10} \left( \mathcal{Z} / \mathcal{Z}_{\rm \Lambda CDM} \right) > 0.5$ ($>1$) is considered substantial (strong) evidence in favor of the model; conversely, a value $<-0.5$ ($<-1$) substantially (strongly) disfavors it; and anything in between $-0.5$ and $0.5$ is barely worth a mention \cite{Kass95}. \refedit{We provide the ``All'' combinations with both DESI DR1+SDSS BAO (which were the values reported in the first submitted version or arXiv v1 of this manuscript) and DESI DR2, to illustrate the precision gain and shifts in parameter values from DR1 to DR2.}
        }
    \renewcommand{\arraystretch}{1.3}
    \begin{ruledtabular}
    \begin{tabular}{lccccc}
    $w_0$ prior & $h$ & $\Omega_{\rm m}$ & $w_0$ & $S_8$ & $\log_{10} \left( \mathcal{Z} / \mathcal{Z}_{\rm \Lambda CDM} \right)$ [$\pm 0.01$] \\
\hline

\textbf{$3 \times 2$pt} \\
$\mathcal{U}(-2, -0.33)$ & -- & $0.348_{-0.044}^{+0.043}$ & $-0.895_{-0.446}^{+0.351}$ & $0.786_{-0.031}^{+0.029}$ & $-0.08$ \\
$\mathcal{U}(-1, -0.33)$ & -- & $0.364_{-0.037}^{+0.039}$ & $-0.694_{-0.206}^{+0.225}$ & $0.799_{-0.023}^{+0.023}$ & $+0.13$ \\
$\delta(-1)$ & -- & $0.339^{+0.037}_{-0.035}$ & -- & $0.782^{+0.020}_{-0.020}$ & -- \\
\textbf{BAO (DESI DR2)} \\
$\mathcal{U}(-2, -0.33)$ & -- & $0.307_{-0.011}^{+0.011}$ & $-0.843_{-0.120}^{+0.112}$ & -- & $-0.36$ \\
$\mathcal{U}(-1, -0.33)$ & -- & $0.308_{-0.010}^{+0.011}$ & $-0.831_{-0.097}^{+0.107}$ & -- & $-0.01$ \\
$\delta(-1)$ & -- & $0.297^{+0.009}_{-0.008}$ & -- & -- & -- \\
\textbf{CMB (no lensing)} \\
$\mathcal{U}(-2, -0.33)$ & $0.741_{-0.097}^{+0.081}$ & $0.263_{-0.050}^{+0.085}$ & $-1.425_{-0.373}^{+0.541}$ & $0.806_{-0.029}^{+0.033}$ & $+0.05$ \\
$\mathcal{U}(-1, -0.33)$ & $0.624_{-0.036}^{+0.029}$ & $0.371_{-0.033}^{+0.047}$ & $-0.759_{-0.172}^{+0.244}$ & $0.844_{-0.020}^{+0.019}$ & $-0.21$ \\
$\delta(-1)$ & $0.666^{+0.009}_{-0.012}$ & $0.326^{+0.016}_{-0.012}$ & -- & $0.830^{+0.017}_{-0.017}$ & -- \\
\textbf{CMB} \\
$\mathcal{U}(-2, -0.33)$ & $0.759_{-0.090}^{+0.070}$ & $0.251_{-0.041}^{+0.072}$ & $-1.506_{-0.311}^{+0.476}$ & $0.803_{-0.024}^{+0.033}$ & $+0.13$ \\
$\mathcal{U}(-1, -0.33)$ & $0.631_{-0.035}^{+0.025}$ & $0.364_{-0.029}^{+0.045}$ & $-0.797_{-0.148}^{+0.234}$ & $0.850_{-0.014}^{+0.015}$ & $-0.31$ \\
$\delta(-1)$ & $0.666^{+0.007}_{-0.010}$ & $0.325^{+0.014}_{-0.010}$ & -- & $0.836^{+0.010}_{-0.010}$ & -- \\
\textbf{SNe Ia} \\
$\mathcal{U}(-2, -0.33)$ & -- & $0.297_{-0.046}^{+0.045}$ & $-0.812_{-0.136}^{+0.118}$ & -- & $-0.24$ \\
$\mathcal{U}(-1, -0.33)$ & -- & $0.292_{-0.044}^{+0.039}$ & $-0.798_{-0.112}^{+0.111}$ & -- & $+0.12$ \\
$\delta(-1)$ & -- & $0.352^{+0.017}_{-0.016}$ & -- & -- & -- \\
\textbf{Strong lensing} \\
$\mathcal{U}(-2, -0.33)$ & $0.740_{-0.077}^{+0.083}$ & $0.341_{-0.149}^{+0.224}$ & $-1.474_{-0.365}^{+0.518}$ & -- & $+0.10$ \\
$\mathcal{U}(-1, -0.33)$ & $0.684_{-0.057}^{+0.064}$ & $0.386_{-0.186}^{+0.263}$ & $-0.766_{-0.170}^{+0.247}$ & -- & $-0.25$ \\
$\delta(-1)$ & $0.712^{+0.066}_{-0.063}$ & $0.345^{+0.248}_{-0.161}$ & -- & -- & -- \\
\textbf{BAO (DESI DR2) + CMB} \\
$\mathcal{U}(-2, -0.33)$ & $0.692_{-0.013}^{+0.014}$ & $0.295_{-0.011}^{+0.011}$ & $-1.053_{-0.084}^{+0.082}$ & $0.813_{-0.008}^{+0.008}$ & $-0.83$ \\
$\mathcal{U}(-1, -0.33)$ & $0.677_{-0.008}^{+0.005}$ & $0.308_{-0.005}^{+0.007}$ & $-0.962_{-0.028}^{+0.048}$ & $0.816_{-0.007}^{+0.007}$ & $-1.02$ \\
$\delta(-1)$ & $0.683^{+0.003}_{-0.003}$ & $0.302^{+0.004}_{-0.004}$ & -- & $0.815^{+0.007}_{-0.007}$ & -- \\
\textbf{BAO (DESI DR2) + SNe Ia} \\
$\mathcal{U}(-2, -0.33)$ & -- & $0.307_{-0.008}^{+0.008}$ & $-0.838_{-0.044}^{+0.044}$ & -- & $+1.55$ \\
$\mathcal{U}(-1, -0.33)$ & -- & $0.307_{-0.007}^{+0.008}$ & $-0.837_{-0.045}^{+0.044}$ & -- & $+1.95$ \\
$\delta(-1)$ & -- & $0.310^{+0.008}_{-0.008}$ & -- & -- & -- \\
\textbf{All (with DESI DR1+SDSS BAO)} \\
$\mathcal{U}(-2, -0.33)$ & $0.667_{-0.006}^{+0.006}$ & $0.318_{-0.006}^{+0.006}$ & $-0.909_{-0.035}^{+0.036}$ & $0.821_{-0.007}^{+0.007}$ & $+0.26$ \\
$\mathcal{U}(-1, -0.33)$ & $0.667_{-0.006}^{+0.006}$ & $0.318_{-0.006}^{+0.006}$ & $-0.908_{-0.035}^{+0.035}$ & $0.821_{-0.007}^{+0.007}$ & $+0.64$ \\
$\delta(-1)$ & $0.680^{+0.003}_{-0.004}$ & $0.307^{+0.005}_{-0.004}$ & -- & $0.821^{+0.007}_{-0.007}$ & -- \\

\textbf{All (with DESI DR2 BAO)} \\
$\mathcal{U}(-2, -0.33)$ & $0.669_{-0.006}^{+0.005}$ & $0.314_{-0.005}^{+0.005}$ & $-0.903_{-0.033}^{+0.034}$ & $0.816_{-0.006}^{+0.006}$ & $+0.56$ \\
$\mathcal{U}(-1, -0.33)$ & $0.669_{-0.006}^{+0.005}$ & $0.314_{-0.005}^{+0.005}$ & $-0.904_{-0.033}^{+0.034}$ & $0.816_{-0.006}^{+0.006}$ & $+0.92$ \\
$\delta(-1)$ & $0.684^{+0.003}_{-0.003}$ & $0.302^{+0.003}_{-0.003}$ & -- & $0.816^{+0.006}_{-0.006}$ & -- \\
\\

	\end{tabular}
    \end{ruledtabular}
    \renewcommand{\arraystretch}{1}
\end{table*}

We can draw a number of conclusions from these results. First, different probes have complementary degeneracies, with SNe nearly orthogonal to BAO and CMB constraints (as is also the case for $w$CDM models with constant $w$). 

Second, the evidence ratios are, in general, sensitive to the adopted prior on $w_0$; this is not surprising, since the $\Lambda$CDM value $w_0=-1$ is at the boundary of the more restrictive NEC prior. On the other hand, in cases where $w_0$ is well constrained by the (combined) data and has a central value greater than $-1$, the resulting value is insensitive to the choice of prior, as expected. 

Third, 
\refedit{\textit{when all the data are combined, we find $w_0=-0.904^{+0.034}_{-0.033}$, about 2.9$\sigma$ discrepant from $\Lambda$CDM, with the evidence ratio favoring the \wphicdm\ model \refedit{substantially (or almost strongly),\footnote{according to the \citet{Kass95} scale}} when the physical NEC prior is imposed.}} \refedit{The parameter shifts from using SDSS+DESI DR1 BAO to DESI DR2 are only 1.3\% for $\Omega_{\rm m}$ and 0.4\% for $w_0$, within (for $\Omega_{\rm m}$) and well within (for $w_0$) the 1$\sigma$ uncertainties.} We note that \refcite{Camilleri24} found $w_0=-0.867^{+0.041}_{-0.040}$ for \wphicdm\,, in 3$\sigma$ disagreement with $\Lambda$CDM, based on DES-SN5YR+SDSS-BAO+CMB \refedit{and using the CMB `shift parameter' in place of a full CMB likelihood analysis. We attribute the shift in $w_0$ to the replacement of SDSS with DESI BAO data, the addition of DES-Y3 3x2pt data, and potentially the different CMB analysis.} 
\refedit{As a check on our results}, we have \refedit{also} verified that our results \refedit{using DESI DR1} are consistent with those of \refcite{Lodha24}, who used the $w_a=-1.58(1+w_0)$ fit, i.e., we find 
$w_0=-0.917\pm 0.035$ when using DES-SN5YR+DESI DR1 BAO+\textit{Planck}+ACT. 

Fourth, \refedit{\textit{the data combination that shows the most significant preference for the \wphicdm\ model is SNe Ia + BAO, with $w_0=-0.837^{+0.044}_{-0.045}$,  $3.6\sigma$ discrepant from $\Lambda$CDM and with the evidence ratio strongly favoring \wphicdm\ regardless of the prior choice.}} 

\refedit{Fifth, as Fig. \ref{fig:omegam_w_constraints} shows, the ``$\Omega_{\rm m}$ tension" in $\Lambda$CDM---the BAO data prefer lower $\Omega_{\rm m}$, CMB is intermediate, and the SN data prefer higher $\Omega_{\rm m}$---is relieved in the \wphicdm\, model with $w_0>-1$ due to the differing parameter degeneracies of the probes.}

Sixth, the best-fit value of $H_0=66.7\pm 0.6$ km s$^{-1}$ Mpc$^{-1}$ for the full data combination (``All" in Table \ref{tab:cosmo_params}) is about 1\% lower than the \textit{Planck} 2018 value for $\Lambda$CDM \cite{PlanckCollaboration18}, slightly exacerbating the tension with low-redshift estimates of the Hubble constant based on the distance ladder \cite{Riess22, Abdalla22}. 

Seventh, as Table \ref{tab:cosmo_params} and Fig.~\ref{fig:s8_constraints} show, in \wphicdm\ with the narrow (NEC) prior on $w_0$, both the large-scale structure and CMB (no lensing) values of $S_8$ and $\Omega_{\rm m}$ shift upward from their $\Lambda$CDM values, increasing the overlap area of the allowed regions in the $S_8$--$\Omega_{\rm m}$ parameter space at 1$\sigma$ when only these two constraints are used. We compute the ``parameter shift'' \cite{Raveri21} as a tension metric \cite{Lemos21} between these two datasets using the \textsc{tensiometer} software package. We find that the two datasets are consistent within 0.9$\sigma$, 1.1$\sigma$, and 1.2$\sigma$ for the \wphicdm\ model with narrow $w_0$ prior, the same model but with the broad $w_0$ prior, and the $\Lambda$CDM model, respectively, indicating a slight reduction in the $S_8$ tension for \wphicdm. However, as Table  \ref{tab:cosmo_params} shows, when all datasets are combined, the resulting value of $S_8$ is insensitive to the model choice, while the central value of $\Omega_{\rm m}$ increases by about 2$\sigma$ in going from $\Lambda$CDM to \wphicdm.

\refedit{Eighth, we can interpret these observational constraints in terms of fundamental parameters of physical models for thawing dark energy. For example, for the massive scalar field model discussed above, the result $w_0=-0.904^{+0.034}_{-0.033}$ implies a scalar field mass $m/H_0 \simeq 0.82\pm 0.13$, where $H_0=2.13h\times 10^{-33}$ eV, an initial value for the scalar field $\phi_{\rm i}/M_{\rm Pl} \simeq 0.55$, and that the field amplitude has changed by about 8\% from its initial value until today. The 95\% upper bound on the scalar mass is approximately $m<1.55\times 10^{-33}$ eV. As discussed in Sec. II, such a small scalar mass is technically natural in pNGB models of ultra-light axions.}

\refedit{Finally, we note that the 2.9$\sigma$ discrepancy between the best-fit \wphicdm\, model and $\Lambda$CDM is smaller than that for the two-parameter $w_0w_a$ model, \refedit{for which} \refcite{DESIDR2} found a 4.2$\sigma$ discrepancy from $\Lambda$CDM for a subset of these datasets (excluding the large-scale structure and strong lensing constraints)} when using wide, flat priors on $w_0$ and $w_a$.
This is to be expected, since the latter parameter space is not constrained to track these thawing models.

\begin{figure}
    \centering
    \includegraphics[width=\columnwidth]{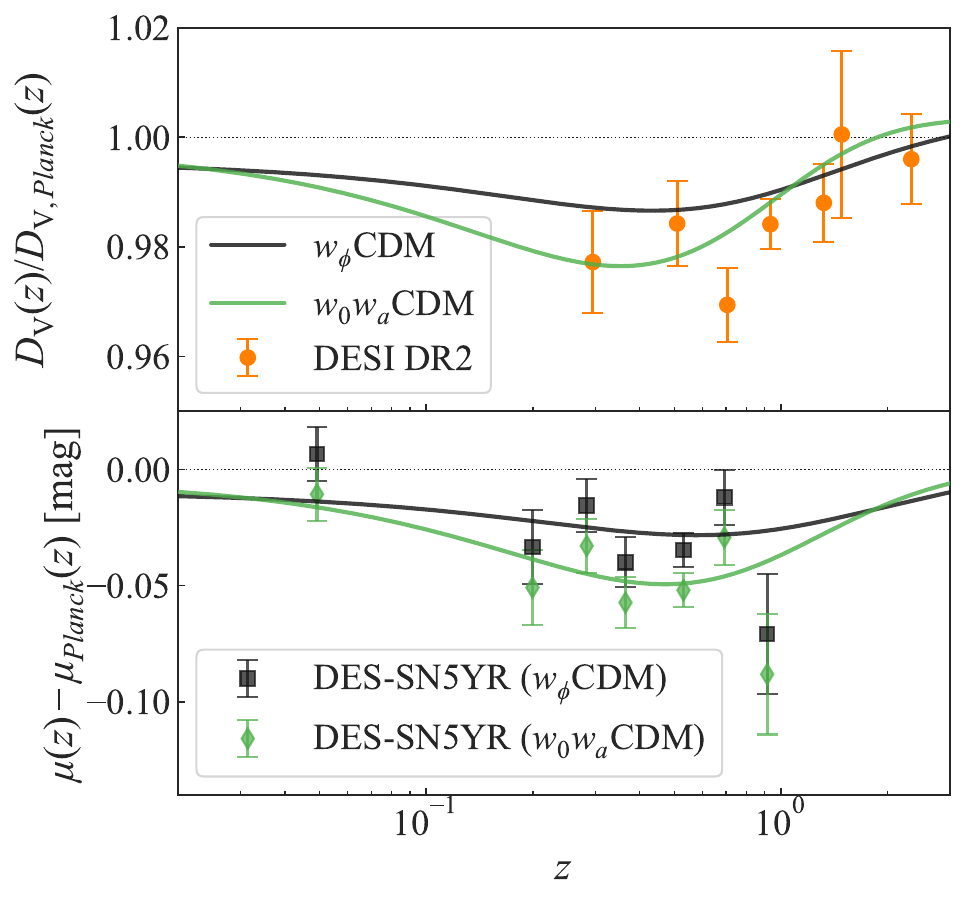}
    \caption{\label{fig:dHres} 
    \refedit{Top panel: amplitude of $D_{\rm V}(z)$ relative to that given by the best-fit $\Lambda$CDM cosmology from \textit{Planck}. The orange points with error bars are from the DESI DR2  measurements. The solid curves show the predicted values using our best-fit \wphicdm\ model (black curve) and the best-fit $w_0w_a$CDM model from \refcite{DESIDR2} (green curve). Bottom panel: distance modulus $\mu(z)$ relative to that given by the best-fit $\Lambda$CDM cosmology from \textit{Planck}. The solid curves represent the predicted values for the same models in the top panel. The data points represent the distance modulus from the DES-SN5YR dataset, as binned by \refcite{DESIDR2} in seven redshift bins. These points can shift uniformly up or down since the absolute magnitude of the SNe is unconstrained. We provide two versions of these points, one that minimizes the error-weighted distance from the best-fit \wphicdm\ model (black points) and one that does so for the best-fit $w_0w_a$CDM model (green points). Subpercent precision in both of these observables in multiple redshift bins up to $z\sim1$ will be necessary to distinguish \wphicdm\ from the $\Lambda$CDM or $w_0w_a$CDM model.}    
    }
\end{figure}

\section{Forecasts}\label{sec:forecasts}

Since the current combined datasets indicate a \refedit{2.9$\sigma$} departure from $\Lambda$CDM for \wphicdm, it is natural to ask how well near-future data will be able to constrain this model and distinguish it from $\Lambda$CDM. The difference in \refedit{$D_{\rm V}(z)$ and distance modulus $\mu(z)$} between our best-fit \wphicdm\ model and the \textit{Planck} 2018 $\Lambda$CDM model \refedit{(and also for the best-fit $w_0w_a$CDM model from \refcite{DESIDR2})} is shown in Fig.~\ref{fig:dHres} and derives largely from the fact that \textit{Planck} precisely fixes the distance to the last scattering surface at $z \sim 1100$. Fig.~\ref{fig:dHres} shows that distinguishing \refedit{\wphicdm\ from the $\Lambda$CDM model (and also from the $w_0w_a$CDM model)} will require \refedit{$D_{\rm V}$ and $\mu(z)$ measurements} with (sub)percent precision in multiple redshift bins to $z \sim 1$. Fortunately, near-future surveys expect to achieve this. Moreover, probes of the growth of structure will provide independent constraints on these models. 

In this section, we provide forecasts for cosmological datasets from imminent or next-generation surveys and facilities. These future datasets are BAO measurements from the extended DESI survey (Section \ref{sec:desi_ext}) and four cosmology probes from the Vera C.~Rubin Observatory Legacy Survey of Space and Time (LSST): $3 \times 2$pt correlations (Section \ref{sec:lsst_3times2}), clusters (Section \ref{sec:lsst_clusters}), SNe Ia (Section \ref{sec:lsst_sn}), and strong lensing (Section \ref{sec:lsst_sl}).  Section \ref{sec:forecast-results} presents the forecast precision on $w_0$ and $\Omega_{\rm m}$ from these future datasets. As the fiducial cosmology for this forecast, we adopt our best-fit \wphicdm\ cosmology with the NEC prior for the \refedit{``All (with DESI DR1+SDSS BAO)''} case in Table \ref{tab:cosmo_params} (i.e., $w_0 = -0.908\pm 0.035$, $\Omega_{\rm m} = 0.821\pm0.007$, and $\alpha = 1.45$).

\subsection{DESI-extension BAO} \label{sec:desi_ext}

The DESI collaboration has proposed an extension of the baseline five-year, 14,000 sq.~deg.~survey that would improve the precision of BAO measurements by increasing the density of Luminous Red Galaxy (LRG) targets, increasing the completeness of emission line galaxy (ELG) targets, and increasing the survey area by 20\% \cite{DESIextension24}. Using their forecast BAO uncertainties, 
we create a mock dataset for $D_{\rm M}/r_{\rm d}$ and $D_{\rm H}/r_{\rm d}$ measurements in 31 redshift bins of width $\Delta z=0.1$ with $0 < z_{\rm eff} < 3.25$. We sample from the corresponding posterior using \textsc{CosmoSIS} and \textsc{emcee} \cite{Foreman-Mackey13}. \refedit{We provide the mock $D_{\rm M}/r_{\rm d}$ and $D_{\rm H}/r_{\rm d}$ measurements we used in Appendix \ref{app:desi_extension}.}

\subsection{LSST $3 \times 2$pt} \label{sec:lsst_3times2}

To forecast constraints from LSST's weak lensing and galaxy clustering measurements, we use the mock Year-10 datasets for galaxy--galaxy, shear--shear, and galaxy--shear two-point correlations generated in the Rubin LSST Dark Energy Science Collaboration (DESC) Science Requirements Document \cite[SRD;][]{LSST18}. The SRD adopted a sky coverage of 14,300 sq.~deg.~after Year 10, which reaches an $i$-band median depth of 26.35 (for 5$\sigma$ point-source detection). The lens sample in the mock dataset adopts ten tomographic bins between $0.2 \leq z \leq 1.2$ with uniform widths of 0.2, with a limiting magnitude of $i_{\rm lim} = 25.3$ and a number density of 48 arcmin$^{-2}$. The source galaxies are divided into five redshift bins, with no upper cutoff in the largest redshift bin, and an effective number density of 27 arcmin$^{-2}$. We forecast the cosmological parameter constraints using the Fisher matrix, adapting the \refedit{DESC-SRD code suite\footnote{\url{https://github.com/CosmoLike/DESC_SRD}} that generates the required mock datasets for this purpose}. The cosmological calculations in this code suite are performed with \textsc{CosmoLike} \cite{Krause17}. We marginalize the effect of calibrateable systematics from the bias in the mean photo-$z$ as implemented in the abovementioned code suite.

\subsection{LSST clusters} \label{sec:lsst_clusters}

We use the mock Year-10 dataset \refedit{generated by the aforementioned DESC-SRD code suite} for the cluster weak lensing and tomographic cluster count analyses. The cluster counts are binned in five redshift bins uniformly spaced between 0.2 and 1 and five richness bins with non-uniform widths between 20 and 220, and no error is assumed in the cluster redshifts. The stacked weak lensing observables are considered in the 1-halo regime. We use the Fisher matrix to obtain the forecast constraints on the cosmological parameters from this dataset, while marginalizing the effect of calibratable systematics on the mean photo-$z$ of the sources.

\subsection{LSST SNe Ia} \label{sec:lsst_sn}

We use the same mock Year-10 SN Ia sample used in the DESC SRD \cite{LSST18}.\footnote{\refedit{The mock dataset including the covariance matrix can be found at \url{https://zenodo.org/records/1409816} \citep{DESCSRDdata}.}} The relative magnitudes of the SNe Ia in this dataset are binned into 40 bins and span the redshift range $0.013 < z < 1.368$. The DESC SRD team inflated the covariance matrix of the $w_0$ and $w_a$ parameters by $\sim$1.5 to account for calibrateable systematics. We find from numerical experiments that this factor of $\sim$1.5 on the $w_0$ -- $w_a$ covariance matrix is similar to multiplying the covariance matrix of the mock data of distance moduli by a factor of $\sim$2. Therefore, we apply this factor of 2 to the covariance matrix of the binned distance moduli before extracting the posterior using \textsc{CosmoSIS} and \textsc{emcee}.

 \subsection{LSST strong lensing} \label{sec:lsst_sl}
 
We consider the same strong lensing dataset used in the forecast by \refcite{Shajib24b}.\footnote{\refedit{We obtain the relevant notebooks and scripts for dataset generation and forecasting from \url{https://github.com/ajshajib/dark_energy_forecast_rubin_strong_lensing}.}} This dataset includes four classes of strong lenses: lensed quasars with time delays, lensed SNe Ia with time delays, double-source-plane lenses, and single-plane lenses with kinematics. The adopted sample sizes for these classes are 236, 153, 87, and 10,000, respectively, which are determined based on forecasts considering lens detectability and achievability of the required follow-up datasets with sufficient signal-to-noise ratio. For the first of these three classes, we use \textsc{hierArc} to simultaneously infer cosmological and population-level galaxy or SN parameters with a hierarchical Bayesian methodology. The mass-sheet degeneracy parameters are allowed to be maximally free in the time-delay cosmography methods \cite{Birrer24, Treu23}, with this degeneracy broken with the kinematics for the lensed quasars (with a subset of 40 systems having JWST-quality resolved kinematics and the rest with unresolved ones) \cite{Shajib18, Yildirim20} and with standardizable magnitudes for the lensed SNe \citep{Birrer22}. We adopt the redshift distribution for the 87 double-source-plane lenses and the uncertainties in the mock data \cite[i.e., the distance ratios $\beta_{\rm s1,s2}$,][]{Collett12} from \refcite{Sharma23}.
 
 \refcite{Li24} provided forecast posteriors from the 10,000 single-plane lenses with kinematics in the $w_0w_a$ parametrization of evolving dark energy. We incorporate the cosmological information from this mock dataset by transforming the posterior in the $w_0w_a$CDM parameter space into a prior in the \wphicdm\ parameter space through a placeholder data vector of angular diameter distances. We construct this data vector with 20 redshift bins between $0.04 \leq z \leq 1.5$, the range spanned by the mock sample of \refcite{Li24}. We first synthesize this data vector using the $w_0w_a$CDM posterior from \refcite{Li24}. We extract the covariance matrix for the bins from the distances sampled from the posterior and retain it to make the exact choice of bin numbers inconsequential. In this step, we inflate the $\Omega_{\rm m}$ and $w_0$ distributions with factors 1.67 and 0.66, respectively, which is a necessary augmentation due to the performed conversion of a non-Gaussian parameter distribution into a data vector with Gaussian covariance; we find these factors by matching the resultant $w$CDM posterior from the synthesized angular diameter distance dataset to that in \refcite{Li24}. After we obtain the covariance matrix of the data vector, we set the mean expansion history encoded by the angular diameter distances to that from our fiducial \wphicdm\ cosmology while also scaling the covariance matrix accordingly. This data vector then effectively provides a prior for the other three classes of strong lenses, with Gaussian standard deviations $\sigma_{\Omega_{\rm m}} = 0.022$, $\sigma_{w_0} = 0.16$, and a Pearson correlation coefficient $r=0.65$. As expected, this is less informative than the $w$CDM posterior (i.e., $\sigma_{\Omega_{\rm m}} = 0.015$, $\sigma_{w_0} = 0.11$, and $r \approx 0$) from \refcite{Li24}, since $w$CDM is a one-parameter extension of $\Lambda$CDM, whereas \wphicdm\ is a quasi-one-parameter one.

\begin{table}
    \caption{
    \label{tab:cosmo_forecast}
        Forecast precision on cosmological parameters from different datasets. The second and third columns provide the 1$\sigma$ uncertainties for $\Omega_{\rm m}$ and $w_0$, respectively. The fourth column provides the Pearson correlation coefficient $r$.}
    \renewcommand{\arraystretch}{1.3}
    \begin{ruledtabular}
    \begin{tabular}{lccc}
    Dataset & $\sigma_{\Omega_{\rm m}}$ & $\sigma_{w_0}$ & $r$ \\
    \hline
DESI-extension BAO & $0.0069$ & $0.077$ & $\phantom{-}0.52$ \\
LSST 3$\times$2pt & $0.0097$ & $0.081$ & $\phantom{-}0.89$ \\
LSST clusters & $0.0148$ & $0.163$ & $\phantom{-}0.82$ \\
LSST SNe Ia & $0.0072$ & $0.024$ & $-0.89$ \\
LSST strong lensing & $0.0118$ & $0.048$ & $-0.19$ \\
Combined & $0.0024$ & $0.011$ & $-0.40$ \\
Combined + CMB (\textit{Planck} + ACT) & $0.0023$ & $0.010$ & $-0.36$ \\
    \end{tabular}
    \end{ruledtabular}
    \renewcommand{\arraystretch}{1}
\end{table}

\begin{figure}
    \centering
    \includegraphics[width=\columnwidth]{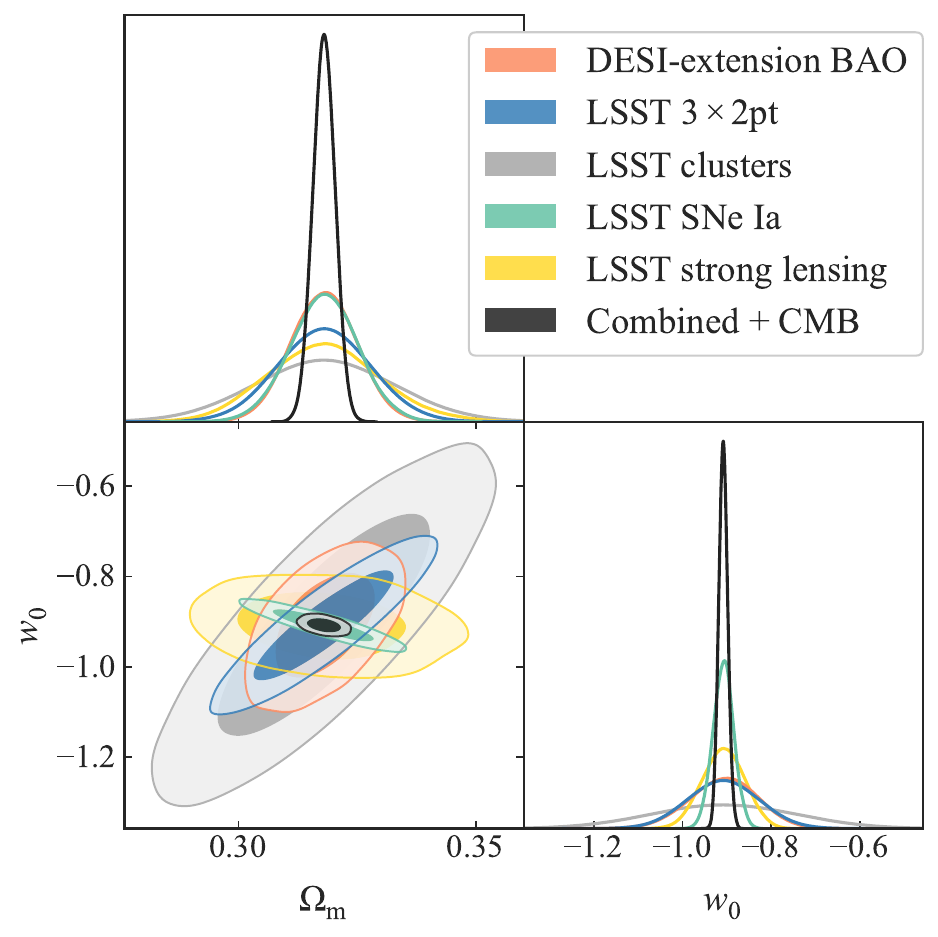}
    \caption{\label{fig:forecast} 
    Forecast constraints on $\Omega_{\rm m}$ and $w_0$ from DESI-extended BAO (orange) and Rubin LSST probes of cosmology after Year 10: $3 \times 2$pt correlations (blue), clusters (grey), SNe Ia (emerald), and strong lensing (yellow). The fiducial values are set at $\Omega_{\rm m} = 0.318$ and $w_0=-0.908$ based on the \refedit{``All (with DESI DR1+SDSS BAO)''} measurement with the NEC prior in Table \ref{tab:cosmo_params}. The Combined $+$ CMB case (black) corresponds to the combination of all the future datasets and the current CMB constraints (from \textit{Planck}$+$ACT, Section \ref{sec:cmb_data}). The displayed posteriors have been adjusted to remove random fluctuations of the means, which are consistent with the statistical uncertainty levels, allowing for clearer visualization of each probe's relative constraining power.
    }
\end{figure}

\subsection{Forecast results} \label{sec:forecast-results}

Table \ref{tab:cosmo_forecast} and Fig.~\ref{fig:forecast} show the forecast constraints on $\Omega_{\rm m}$ and $w_0$ for the \wphicdm\ model from these near-future datasets and their combination. These numbers should be interpreted with caution, since they depend on assumptions about system performance, systematic error levels, and nuisance parameters that often are not well determined ahead of time. However, we expect the broad trends shown here to be robust: the probes sensitive to the growth of structure ($3\times2$pt and clusters) show the strongest correlation between $\Omega_{\rm m}$ and $w_0$, followed by BAO. Conversely, SNe show an anti-correlation as in the current data, and strong lensing shows a very weak anti-correlation. These complementary degeneracies mean that the combined constraints will be very tight, and the fact that we have multiple probes with similar degeneracies will provide systematic cross-checks between them.

The combined forecast constraints on each of the two parameters are about three times stronger than the current combined constraints. We also note that these constraints are strong enough that including current CMB data yields only very modest improvements. If the combined central value of $w_0$ were to remain at $-0.908$, it would go from a 2.6$\sigma$ deviation from $\Lambda$CDM to a 9.2$\sigma$ one. Therefore, the prospects for testing physical models of evolving dark energy models as the source of the recent SN+BAO+CMB results appear quite good.

Finally, as noted above, current data place no meaningful constraints on the evolution parameter $\alpha$ of Eq. \ref{eq:walpha}, so it is interesting to ask whether future data might do so. We have carried out forecasts using the LSST SN Ia dataset for the same fiducial \wphicdm\ model, but allowing a very broad, flat prior on $\alpha$ in the fit (with fiducial value $\alpha = 1.45$), and find only weak upper bounds on $\alpha$ of 9.8, 18.5, and 59 at 84th, 90th, and 95th percentiles, respectively. In this case with the broad $\alpha$ prior, there is also a degeneracy between the inferred values of $\alpha$ and $\Omega_{\rm m}$ that leads to bias in the marginalized value of $\Omega_{\rm m}$. This imperfect degeneracy can be interpreted as that the expansion history required by the data can be retained with a simultaneous increase in $\alpha$ and $\Omega_{\rm m}$, where their impact on changing the epoch of matter--dark energy equality counteracts each other at fixed or well-constrained $w_0$.

\section{Conclusion}\label{sec:conclusion}

We have derived current and forecast near-future parameter constraints on a broad class of simple, physically well-motivated, evolving scalar-field dark energy models, using a quasi-one-dimensional parametrization for the evolution of the scalar-field equation of state parameter, $w_\phi(z)$, that is surprisingly accurate for these models. Within this class of models, the current data prefer (at \refedit{2.9$\sigma$}) a model with scalar-field mass just slightly below the current value of the Hubble parameter, \refedit{$m = (0.82\pm 0.13)H_0=(1.2\pm 0.2)\times 10^{-33}$ eV} (for a free, massive scalar), corresponding to a current value of the equation of state parameter \refedit{$w_0=-0.904^{+0.034}_{-0.033}$}, with a Bayesian evidence ratio substantially in favor of these models over $\Lambda$CDM when using a physical (NEC-constrained) prior on $w_0$.

While there are scalar field models that provide a better fit to the current data than the class of models we have considered \cite{Shlivko24}, for the reasons outlined in Sec. \ref{sec:theory}, we have chosen to impose a theory prior that excludes such behavior.

For the class of models we have considered, near-future data from the extended DESI survey and the Rubin Observatory LSST should improve upon current constraints on $\Omega_{\rm m}$ and $w_0$ by a factor of three each, in which case a model with the current central value of $w_0$ would be distinguishable from $\Lambda$CDM with very high confidence ($9.2\sigma$) and convincing Bayesian evidence. This is an exciting prospect.

\begin{acknowledgments}
The authors thank Shrihan Agarwal, Dhayaa Anbajagane, Austin Joyce, Tanvi Karwal, Qinxun Li, Yuuki Omori, David Shlivko, Michael Turner, Georgios Valogiannis, and Joe Zuntz for helpful discussions. The authors also thank Simon Birrer, Martin Millon, and Judit Prat for having implemented the TDCOSMO likelihood in \textsc{CosmoSIS} at the ``Lensing at Difference Scales'' workshop hosted by KICP in 2023. Support for this work was provided by NASA through the NASA Hubble Fellowship grant HST-HF2-51492 awarded to AJS by the Space Telescope Science Institute (STScI), which is operated by the Association of Universities for Research in Astronomy, Inc., for NASA, under contract NAS5-26555. AJS also received support from NASA through STScI grants HST-GO-16773 and JWST-GO-2974.

This work was completed in part with resources provided by the University of Chicago’s Research Computing Center. This paper made use of \href{https://github.com/joezuntz/cosmosis}{\textsc{CosmoSIS}} \citep{Zuntz15}, \textsc{tensiometer} \cite{Raveri21}, \textsc{emcee} \cite{Foreman-Mackey13}, \textsc{schwimmbad} \cite{Price-Whelan17}, \textsc{GetDist} \cite{Lewis19}, \textsc{numpy} \cite{Oliphant15}, \textsc{astropy} \cite{AstropyCollaboration13,AstropyCollaboration18,AstropyCollaboration22}, \textsc{scipy} \cite{Jones01}, \textsc{mpi4py} \cite{Dalcin05, Dalcin08, Dalcin11, Dalcin21,Rogowski23}, \textsc{jupyter} \cite{Kluyver16}, \href{https://github.com/lenstronomy/lenstronomy}{\textsc{lenstronomy}} \cite{Birrer18, Birrer21b}, \href{https://github.com/sibirrer/hierArc}{\textsc{hierArc}}, \textsc{PArthENoPE} \cite{Pisanti08}, and \textsc{fast-pt} \cite{McEwen16}.	
\end{acknowledgments}

\section*{Data Availability}

The data that support the findings of this article are openly available \cite{DESI24, DESIDR2, PlanckCollaboration20, Madhavacheril24, Abbott22, DESCollaboration24, Birrer20, LSST18, DESCSRDdata}.


\bibliography{ajshajib}

\appendix
\section{DESI-extension mock BAO dataset} \label{app:desi_extension}

\refedit{In this appendix, we provide the mock BAO measurements of $D_{\rm M}/r_{\rm d}$ and $D_{\rm H}/r_{\rm d}$ in 31 redshift bins in Table \ref{tab:desi_extended}. These mock values were synthesized for a fiducial \wphicdm\ cosmology with $\Omega_{\rm m} = 0.318$, $w_0 = -0.908$, and $\alpha=1.45$. }

\begin{table}
    \caption{
    \label{tab:desi_extended}
        \refedit{Mock BAO measurements for the DESI-extension survey.}}
    \renewcommand{\arraystretch}{1.3}
    \begin{ruledtabular}
    \begin{tabular}{lccc}
    Tracer & $z_{\rm eff}$ & $D_{\rm M}/r_{\rm d}$ & $D_{\rm H}/r_{\rm d}$ \\
    \hline
\texttt{BGS1} & 0.05 & $\phantom{0}1.524 \pm 0.089$ & $24.38 \pm 3.75$ \\
\texttt{BGS2} & 0.15 & $\phantom{0}4.332 \pm 0.100$ & $26.47 \pm 1.26$ \\
\texttt{BGS3} & 0.25 & $\phantom{0}6.979 \pm 0.102$ & $26.22 \pm 0.81$ \\
\texttt{BGS4} & 0.35 & $\phantom{0}9.643 \pm 0.119$ & $24.78 \pm 0.59$ \\
\texttt{LRG1} & 0.45 & $11.802 \pm 0.121$ & $23.63 \pm 0.46$ \\
\texttt{LRG2} & 0.55 & $14.033 \pm 0.121$ & $22.14 \pm 0.37$ \\
\texttt{LRG3} & 0.65 & $16.205 \pm 0.120$ & $20.48 \pm 0.32$ \\
\texttt{LRG4} & 0.75 & $18.279 \pm 0.132$ & $19.03 \pm 0.26$ \\
\texttt{LRG5} & 0.85 & $19.904 \pm 0.131$ & $17.96 \pm 0.22$ \\
\texttt{LRG6} & 0.95 & $21.628 \pm 0.151$ & $16.59 \pm 0.22$ \\
\texttt{LRG7} & 1.05 & $23.580 \pm 0.217$ & $16.33 \pm 0.25$ \\
\texttt{ELG1} & 1.15 & $25.108 \pm 0.260$ & $15.01 \pm 0.23$ \\
\texttt{ELG2} & 1.25 & $26.298 \pm 0.281$ & $14.60 \pm 0.22$ \\
\texttt{ELG3} & 1.35 & $28.265 \pm 0.303$ & $13.56 \pm 0.21$ \\
\texttt{ELG4} & 1.45 & $29.686 \pm 0.331$ & $12.77 \pm 0.21$ \\
\texttt{ELG5} & 1.55 & $30.387 \pm 0.463$ & $12.00 \pm 0.24$ \\
\texttt{QSO1} & 1.65 & $30.066 \pm 0.978$ & $11.38 \pm 0.52$ \\
\texttt{QSO2} & 1.75 & $32.109 \pm 1.001$ & $11.00 \pm 0.51$ \\
\texttt{QSO3} & 1.85 & $35.084 \pm 1.119$ & $10.51 \pm 0.53$ \\
\texttt{QSO4} & 1.95 & $35.707 \pm 1.275$ & $\phantom{0}9.89 \pm 0.48$ \\
\texttt{QSO5} & 2.05 & $36.656 \pm 1.337$ & $10.15 \pm 0.51$ \\
\texttt{Ly$\alpha$1} & 2.15 & $36.949 \pm 0.669$ & $\phantom{0}9.48 \pm 0.18$ \\
\texttt{Ly$\alpha$2} & 2.25 & $38.334 \pm 0.749$ & $\phantom{0}8.76 \pm 0.18$ \\
\texttt{Ly$\alpha$3} & 2.35 & $38.909 \pm 0.825$ & $\phantom{0}8.01 \pm 0.17$ \\
\texttt{Ly$\alpha$4} & 2.45 & $38.708 \pm 0.909$ & $\phantom{0}8.40 \pm 0.20$ \\
\texttt{Ly$\alpha$5} & 2.55 & $41.338 \pm 1.047$ & $\phantom{0}8.09 \pm 0.20$ \\
\texttt{Ly$\alpha$6} & 2.65 & $41.941 \pm 1.281$ & $\phantom{0}7.42 \pm 0.21$ \\
\texttt{Ly$\alpha$7} & 2.75 & $41.870 \pm 1.441$ & $\phantom{0}7.18 \pm 0.22$ \\
\texttt{Ly$\alpha$8} & 2.85 & $41.696 \pm 1.859$ & $\phantom{0}7.08 \pm 0.26$ \\
\texttt{Ly$\alpha$9} & 2.95 & $42.184 \pm 2.145$ & $\phantom{0}6.69 \pm 0.28$ \\
\texttt{Ly$\alpha$10} & 3.25 & $46.051 \pm 1.827$ & $\phantom{0}5.81 \pm 0.19$ \\
    \end{tabular}
    \end{ruledtabular}
    \renewcommand{\arraystretch}{1}
\end{table}

\end{document}